# THE VERY-SOFT X-RAY EMISSION
# OF X-RAY FAINT EARLY-TYPE GALAXIES

S. PELLEGRINI[1,2] AND G. FABBIANO[3]

[1] European Southern Observatory, Karl Schwarzschild str.2,
   D-85748 Garching bei München
[2] Dipartimento di Astronomia, Università di Bologna,
   via Zamboni 33, I-40126 Bologna
[3] Harvard-Smithsonian Center for Astrophysics, 60 Garden str.,
   Cambridge, MA 02138

# ABSTRACT

A recent re-analysis of *Einstein* data, and new *ROSAT* observations, have revealed the presence of at least two components in the X-ray spectra of X-ray faint early-type galaxies: a relatively hard component ($kT > 1.5$ keV), and a very soft component ($kT \sim 0.2-0.3$ keV). In this paper we address the problem of the nature of the very soft component, and whether it can be due to a hot interstellar medium (ISM), or is most likely originated by the collective emission of very soft stellar sources. To this purpose, hydrodynamical evolutionary sequences for the secular behavior of gas flows in ellipticals have been performed, varying the type Ia supernovae rate of explosion, and the dark matter amount and distribution. The results are compared with the observational X-ray data: the average *Einstein* spectrum for 6 X-ray faint early-type galaxies (among which are NGC4365 and NGC4697), and the spectrum obtained by the *ROSAT* pointed observation of NGC4365. The very soft component could be entirely explained with a hot ISM only in galaxies such as NGC4697, i.e. when the depth of the potential well – on which the average ISM temperature strongly depends – is quite shallow; in NGC4365 a diffuse hot ISM would have a temperature larger than that of the very soft component, because of the deeper potential well. So, in NGC4365 the softest contribution to the X-ray emission comes certainly from stellar sources. As stellar soft X-ray emitters, we consider late-type stellar coronae, supersoft sources such as those discovered by *ROSAT* in the Magellanic Clouds and M31, and RS CVn systems. All these candidates can be substantial contributors to the very soft emission, though none of them, taken separately, plausibly accounts entirely for its properties.

We finally present a model for the X-ray emission of NGC4365, to reproduce in detail the results of the *ROSAT* pointed observation, including PSPC spectrum and radial surface brightness distribution. The present data may suggest that the X-ray surface brightness is more extended than the optical profile. In this case, a straightforward explanation in terms of stellar sources could be not satisfactory. The available data can be better explained with three different contributions: a very soft component of stellar origin, a hard component from X-ray binaries, and a $\sim 0.6$ keV hot ISM. The latter can explain the extended X-ray surface brightness profile, if the galaxy has a dark-to-luminous mass ratio of 9, with the dark matter very broadly distributed, and a SNIa explosion rate of $\sim 0.6$ the Tamman rate.

# 1. Introduction

After the launch of the *Einstein* satellite (Giacconi *et al.* 1979), it was first realized through X-ray measurements that elliptical galaxies may retain a large amount ($\sim 10^8 - 10^{11} M_\odot$) of hot ($T \sim 10^7$ K) interstellar gas (e.g. Forman, Jones, and Tucker 1985; Canizares, Fabbiano, and Trinchieri 1987; see the review of Fabbiano 1989). This conclusion was supported by the presence of X-ray emission displaced from the optical body in some cases, by the relatively soft ($\sim$1 keV) X-ray spectra of X-ray bright ellipticals, and by the steep correlation between X-ray and optical luminosity ($L_X \propto L_B^{1.5-2}$), see Fig. 1. In contrast in spiral galaxies, where the X-ray emission can be attributed to a population of evolved stellar sources (see Fabbiano 1989, and references therein), the X-ray spectra tend to be harder, and $L_X \propto L_B$. Besides suggesting a steep relationship between $L_X$ and $L_B$, the $L_X$–$L_B$ diagram of early-type galaxies shows a substantial amount of scatter: the range of $L_X$ for a given optical luminosity spans about two orders of magnitude, for $L_B > 3 \times 10^{10} L_\odot$ (Fig. 1).

Many theoretical models were developed to explain the $L_X$–$L_B$ diagram, including numerical simulations for the behavior of gas flows fed by stellar mass loss and heated to X-ray temperatures by type Ia supernovae (SNIa; see the review of Fabbiano 1989; Loewenstein and Mathews 1987; Sarazin and White 1987, 1988; D'Ercole *et al.* 1989; David *et al.* 1991; Ciotti *et al.* 1991, hereafter CDPR). The most successful models are those explaining the scatter in the $L_X$–$L_B$ diagram in terms of different dynamical phases for the hot gas flows, ranging from winds to inflows (CDPR). These authors, assuming that the SNIa explosion rate is declining with time slightly faster than the rate with which mass is lost by stars, find that in the beginning the energy released by SNIa's can drive the gas out of the galaxies through a supersonic wind. As the collective SNIa's energy input decreases, a subsonic outflow takes place, and this gradually slows down until a central cooling catastrophe leads to the onset of an inflow. An attractive feature of this scenario is that any of the three phases wind/outflow/inflow can be found at the present epoch, depending only on the various depths and shapes of the potential wells of the galaxies: in X-ray faint ellipticals the gas is still in the wind phase, and the emission is mostly accounted for by stellar sources; in the bulk of galaxies the gas flows are in the outflow phase; in X-ray brightest galaxies the soft X-ray emitting gas dominates the emission, being in the inflow phase. In this latter case, the gas flow resembles a "cooling flow" (e.g. Sarazin and White 1987, 1988).

The knowledge of the actual contribution of discrete sources to the total X-ray emission is an important tool for constraining the gas flow phase, especially in X-ray faint galaxies. Unfortunately, even the amount of the hard contribution coming from accreting low-mass binaries, which should be substantial, is still a matter of debate, due to the impossibility of estimating it directly from the available *Einstein* X-ray spectra (Forman, Jones, and Tucker 1985; Canizares, Fabbiano, and Trinchieri 1987). Progress in this field comes from a recent systematic investigation of the X-ray spectra of early-type galaxies obtained by the Imaging Proportional Counter (IPC) on board of the *Einstein* satellite, by Kim, Fabbiano, and Trinchieri 1992 (see Fig. 1). These authors find that on the average the X-ray emission temperature increases with decreasing X-ray to optical ratio ($L_X/L_B$), until the dominant contribution to the total emission comes from a hard component, similar to that dominating the emission in spirals, which can be fitted with a $kT > 3$ keV bremsstrahlung model. These findings are in agreement with the CDPR scenario, in which the emission of X-ray faint early-type galaxies is due mainly to a stellar component (these galaxies would be in the wind phase), while the gaseous emission becomes dominant in X-ray bright galaxies, which would be in the inflow phase.

In those galaxies with the lowest values for $L_X/L_B$ (hereafter Group 1, which includes 6 early-type galaxies with $29.3 < \log(L_X/L_B) < 30.0$, with $L_X$ in erg s$^{-1}$, $L_B$ in $L_\odot$), Kim, Fabbiano, and Trinchieri 1992 find that a very soft ($kT \approx 0.2$ keV) thermal X-ray component (hereafter VSC) is present in addition to the hard thermal component. The VSC emission has been estimated to amount approximately to one third to half ($\sim 10^{40}$ erg s$^{-1}$) of the total X-ray luminosity in the 0.2–3.5 keV band. With the available data, it remains possible that this VSC is present in all early-type galaxies. Recent *ROSAT* PSPC observations of two galaxies of Group 1 (NGC4365 and NGC4382, Fabbiano *et al.* 1994, the companion paper) confirm the existence of a VSC.

Investigating the nature of the VSC is the principal aim of this paper. We present hydrodynamical evolutionary sequences for the secular behavior of gas flows – in the frame of the CDPR models – specific to the Group 1 galaxies NGC4697 and NGC4365, to explore whether such flows can account for the observed VSC characteristics. These two galaxies were chosen because of their different central stellar velocity dispersions $\sigma_*$; $\sigma_*^2$ is proportional to the central depth of the galaxy potential well, on which the gas temperature strongly depends. In our models we adopt the observed optical characteristics of the galaxies, and we vary the SNIa's rate of explosion, and the amount and distribution of dark matter. The *ROSAT* results on NGC4365 allow us to make more secure conjectures on the various components of the X-ray emission, considering both the X-ray spectrum and the X-ray surface brightness profile. We also consider whether the VSC could originate in the collective emission of discrete very soft sources, such as late-type stellar coronae, supersoft accreting binaries like those detected by *ROSAT* in the Magellanic Clouds and M31 (Greiner *et al.* 1991; Kahabka and Pietsch 1992; Schaeidt *et al.* 1993), and RS CVn systems.

This paper is organized as follows: in §2 we summarize and re-examine the observational properties of the VSC in X-ray faint early-type galaxies; in §3 we investigate whether these properties are consistent with the expected characteristics of gas flows in Group 1 galaxies; in §4 we show the results of the numerical simulations to model the observations of NGC4697 and NGC4365; in §5 we consider stellar sources as origin of the VSC; in §6 we present a detailed model for the X-ray emission of NGC4365; in §7 we summarize our conclusions.

## 2. The VSC from *Einstein* and *ROSAT* Observations

For an accurate theoretical investigation, we need to have the best definition of the VSC characteristics which is possible within the observational boundaries. So, our first step was to re-analyze in detail the average spectrum obtained from *Einstein* IPC data for Group 1 galaxies by Kim, Fabbiano, and Trinchieri 1992, to determine the range of the VSC temperature ($T_{VSC}$) and the VSC contribution to the total X-ray emission ($L_{X,VSC}$) allowed by the data. To this purpose, we used the PROS ('xray') package of *IRAF*, developed at *SAO* for the analysis of reduced X-ray data. Following Kim, Fabbiano, and Trinchieri 1992, we adopt an average value for the hydrogen column density along the line of sight $N_H = 3.2 \times 10^{20}$ cm$^{-2}$. A one-temperature optically thin thermal model with solar abundance gives an unacceptable best-fit $\chi^2$ (minimum $\chi^2 = 17.5$ for 7 degrees of freedom). Acceptable fits are obtained when the spectrum is modeled by the superposition of a hard thermal component ($kT > 1$ keV), and a soft thermal component ($kT < 0.5$ keV), for solar abundance of the emitting gas; in Fig. 2 the best fit case is represented (minimum $\chi^2 = 7.9$ for 5 degrees of freedom)[1]. The ranges of values given above for the temperatures correspond to the 90% confidence level, for three interesting parameters: the two temperatures, and the ratio of the normalizations of the two thermal components (Fig. 3). The temperatures are not well constrained for very low and high values, because the combined mirror and IPC spectral response drops significantly below 0.5 keV and beyond 4 keV, and the IPC has very poor spectral resolution.

The uncertainties in Group 1 data are such that the normalizations of the two thermal components are not well determined. As an example, if we keep $1 < kT < 5$ for the hard component, and $0.1 < kT_{VSC} < 0.5$ (see Fig. 3), we have that $0.15 < L_{X,VSC}/L_X < 0.50$ at the 90% confidence level. Assuming that this range is representative of all Group 1 galaxies, we derive the range of $L_{X,VSC}$ values given in Table 1 for NGC4697.

Recent *ROSAT* observations of two Group 1 galaxies (NGC4382 and NGC4365) confirm the presence of a VSC with better spectral data. We do not describe these observations here, because they are discussed in detail in the companion paper (Fabbiano *et al.* 1994). Again a one-temperature model is excluded, except for the case of a temperature between 0.6 and 1.1 keV, and abundance values close to zero. Another more realistic possibility (Fabbiano *et al.* 1994) is the combination of a hard ($kT > 1.5$ keV) and a soft thermal component, with $kT_{VSC} = 0.2$ keV at the best fit. In Fig. 4a we show the ranges of variation at 68%, 90% and 99% confidence level for $kT_{VSC}$ in the NGC4365 case (line of sight absorption has been assumed, see Table 1). A comparison with the similar figure obtained from the

---

[1] With these data we cannot model temperatures and metal abundance of the ISM meaningfully. However the *ROSAT* observations of two Group 1 galaxies (Fabbiano *et al.* 1994) suggest that the two temperature model is the more likely representation of the data.

*Einstein* IPC data (Fig. 3), shows how the contours in Fig. 4a are still open towards high temperatures, due to the PSPC spectral response, but now the VSC temperature is well constrained. The soft and the hard contributions to the total X-ray luminosity in the *ROSAT* band are comparable in the best fit case, but can vary up to a factor of two in each direction (Fabbiano *et al.* 1994). The spectral counts distribution of NGC4365 together with the two best fit components and their sum are shown in Fig. 4b. We report in Table 1 the main results of the analysis concerning NGC4365, for which we perform numerical simulations of the hot gas behavior (we do not simulate the S0 galaxy NGC4382, because our spherically symmetric models cannot handle the presence of a disk).

### 3. Can a hot ISM be the origin of the VSC?

In this section we examine whether the observed VSC characteristics are in principle compatible with a hot ISM origin. As an outcome of this investigation, we will also find what input parameters are needed for the numerical simulations of the gas flow behavior. We first introduce our galaxy models.

*3.1 The galaxy models*

The galaxy structure adopted for the hydrodynamical simulations of the hot gas evolution is that of CDPR: it is a superposition of a King 1972 model for the luminous matter distribution, and of a quasi-isothermal halo for that of the dark matter; both distributions are truncated at the same tidal radius $r_t$. We adopt $r_t/r_{c*}=180$, where $r_{c*}$ is the stellar core radius, so that the projection of the luminous matter distribution gives a good fit to the observed optical surface brightness profiles of ellipticals (Kormendy 1982). The observed quantities are $L_B$, $\sigma_*$, and $r_{c*}$; from these the central stellar density $\rho_{0*}$ is derived through the virial condition (e.g. Binney and Tremaine 1987); in this way the luminous matter distribution is completely defined, because the stellar mass-to-light ratio is assumed to be constant with radius. The dark matter distribution can be determined by choosing the ratios of the dark/luminous masses ($R = M_h/M_*$), the dark/luminous core radii ($\beta = r_{ch}/r_{c*}$), and the dark/luminous central densities ($\gamma = \rho_{0h}/\rho_{0*}$), only two among these three parameters being independent. The problem of the physical consistency of the proposed mass distribution – whether it represents a self-consistent solution of the Poisson and Vlasov equations (see, e.g., Binney and Tremaine 1987) with two components – was addressed by Ciotti and Pellegrini 1992, in the hypothesis of stellar orbits with various anisotropies. In this paper we just want to derive the physical properties of the gas flows, and possible constraints on the amount and distribution of the dark matter, starting from reasonable shapes for the potential wells.

The time-evolving input ingredients of the numerical simulations are the rate of stellar mass losses, and the SNIa's heating; these too are calculated as in CDPR. In particular, the stellar mass loss rate is parameterized as $\dot{M}_*(t) \simeq -1.5 \times$ $10^{-11} L_{\rm B} (t/15 \,{\rm Gyrs})^{-1.3}$ $M_\odot$ yr$^{-1}$; the SNIa's heating is parametrized as $L_{SN}(t) \simeq$ $7.1 \times 10^{30} \vartheta_{\rm SN} L_{\rm B}(t/15 \,{\rm Gyrs})^{-1.5}$ erg s$^{-1}$, with $0 < \vartheta_{\rm SN} < 1$. The value $\vartheta_{\rm SN} = 1$ gives the present rate of SNIa explosion in ellipticals estimated by Tammann 1982; this is $0.88 h^2$SNU, where $h = H_\circ/100$, and 1 SNU=1 event/100 yrs/$10^{10}$ $L_\odot$ (we adopt $H_\circ = 50$ km/sec/Mpc throughout this paper). Van den Bergh and Tammann 1991 give an estimate which translates into $\vartheta_{\rm SN} = 1.1$, close to that of Tammann 1982; Cappellaro *et al.* 1993 give a value 4 times lower ($\vartheta_{\rm SN} \sim 0.3$), which illustrates the present range of uncertainty in the SNIa rate of explosion in ellipticals. $L_{SN}$ is assumed here to be decreasing with time just faster than the stellar mass loss rate (Renzini 1989). This assumption produces a time decreasing specific heating of the gas, and so the wind/outflow/inflow secular evolution of the gas dynamical state, described in §1.

The effects introduced by the possible presence of population gradients can be neglected without affecting our conclusions. Age and/or metallicity gradients – such as those that can be estimated from observations – introduce a variation of $\dot{M}_*(t)$ within $\sim 10\%$ (Renzini and Buzzoni 1986), while many uncertainties already affect $L_{SN}(t)$.

In summary, our parameter space for each galaxy model consists of $(R, \beta, \vartheta_{\rm SN})$, the first two parameters describing the dark matter amount and concentration, the third describing the present rate of SNIa explosions.

### 3.2 $L_{\rm X,gas}$ and $L_{\rm X,VSC}$

Here we address the question of which dynamical phase of the gas flow can reproduce the observed $L_{\rm X,VSC}$. Global inflows are far too luminous ($L_{\rm X,gas} \gtrsim 10^{41}$ erg s$^{-1}$) with respect to $L_{\rm X,VSC}$, while during winds the X-ray luminosity of the gas is very low ($< 10^{39}$ erg s$^{-1}$); during the intermediate outflow phase $L_{\rm X,gas}$ can vary by orders of magnitude, going from wind values to inflow values. The duration of this transient phase is very sensitive to the input parameters, and it can last up to $\sim 10$ Gyrs; this means that any of the $L_{\rm X}$ values, from those typical of winds to those typical of inflows, can be shown by gas flows in outflows *at the present epoch*, depending on the input parameters. Another dynamical state in which $L_{\rm X,gas} \sim L_{\rm X,VSC}$ is the "partial wind" (hereafter PW). In this case radiative losses suppress an outflow in the center of the galaxy, causing a small inflow region, but a wind can be sustained in the external regions, where the gas density is much lower and the gas is also less tightly bound (MacDonald and Bailey 1981; §4.1.1). Clearly, the smaller the central inflow region, and the higher the wind velocity, the lower is the gas mass inside the galaxy, and so the X-ray luminosity.

Which gas flow phase is expected in Group 1 galaxies? They show a large range of values for the central stellar velocity dispersion $\sigma_*$ ($\sigma_* = 186$–$284$ km s$^{-1}$; Whitmore, McElroy, and Tonry 1985), reaching particularly low values with respect to the average $L_{\rm B}$–$\sigma_*$ relation for early-type galaxies (Faber & Jackson 1976, Davies *et al.* 1983). This is important because $\sigma_*$ is one of the key parameters regulating the gas flow phase; together with the amount of dark matter ($R$), and its concentration ($\beta$), it measures the binding energy of the gas. Following CD-PR, we can summarize the gas flow energetic balance as a function of $\sigma_*$, and the introduced parameters ($R$, $\beta$, $\vartheta_{\rm SN}$), by using the parameter $\chi$. This is defined as the ratio of the power required to steadily extract the gas shed by stars from the potential well of the galaxy, to the energy made available by SNIa's explosions: $\chi \propto \sigma_*^2(A + \beta^{-0.14}R)/\vartheta_{\rm SN}$, at any epoch ($A$ is a constant). When $\chi > 1$, inflows are expected, while when $\chi < 1$, we expect winds; if $\chi \approx 1$ we have outflows. In Table 2 we show the values assumed by $\chi$ at the present epoch, for $\vartheta_{\rm SN} = 1$ and for the $\sigma_*$ values of the two galaxies in Table 1, NGC4697 and NGC4365; the dark matter amount ($R$) and concentration ($\beta$) have been varied. We can see how winds are expected in both galaxies, if $\vartheta_{\rm SN} = 1$, even for large $R$ values. So, we have to lower $\vartheta_{\rm SN}$ to obtain outflows, especially for NGC4697, which has the smaller $\sigma_*$.

The parameter $\chi$ cannot be used to make predictions on the PW phase: the gas could be in PW with $\chi$ either larger or smaller than 1, because this parameter describes the global energetic balance of the gas, while in PW the gas behaves differently in the center and in the outskirt of the galaxy (these regions become "decoupled"). We expect PW solutions for small amounts of dark matter – since we assume it is distributed mainly at large radii – and this is confirmed by the numerical simulations (see also §4.1.1 and 4.2.1).

In conclusion, we have to explore through numerical simulations of the gas flow evolution which region of the parameter space ($R$, $\beta$, $\vartheta_{\rm SN}$) is populated by gas flow phases with $L_{\rm X,gas} \sim L_{\rm X,VSC}$, at the present epoch. We can just anticipate we need $\vartheta_{\rm SN} < 1$.

### 3.3 $T_{gas}$ and $T_{VSC}$

We explore here whether the VSC temperature is compatible with that of the hot ISM in ellipticals. As it was the case for the X-ray luminosity, the average gas flow temperature depends strongly on $\sigma_*$. For instance, during the inflow phase predictions on the temperature profile of the gas can be made where the following simplifying assumptions hold: 1) the gas is in quasi-hydrostatic equilibrium in the galaxy potential well (i.e. $t_{sound}(r) << t_{cool}(r)$, with $t_{sound}$ the sound speed crossing time, and $t_{cool}$ the gas cooling time); 2) the flow is dominated by gravity rather than by the supernovae energy input (the gas is in inflow); 3) the cooling is just a small fraction of the global energetic balance of the gas flow. Under these hypotheses, one derives for the gas $T_{in}(r) \propto |\phi(r)|$, $\phi$ being the total galaxy gravitational potential. Just outside the central region of the galaxy, as soon as the infall velocity becomes small and the above hypotheses apply, one has $kT_{in}(5r_{c*}) \approx 3.6\mu m_P \sigma_*^2 F(R,\beta)$, where $\mu m_P$ is the average gas particle mass, and $F(R,\beta)$ is a form factor of the order of unity; $F(R,\beta) = 0.46 \div 1.6$, as $R = 0 \div 10$, and $\beta = 2$. So $kT_{in}(5r_{c*}) \gtrsim 0.3$ keV if $\sigma_* \gtrsim 180$ km s$^{-1}$.

In conclusion, we can expect the gas flows to be characterized by temperatures of the order of $T_{VSC}$ only for low $\sigma_*$ values. However the temperature estimates given above refer to the central regions of the galaxy, while the gas radial temperature distribution decreases outward, unless a sizable external pressure is effective, or a very large amount of dark matter broadly distributed (see, e.g., Sarazin & White 1987). So, a detailed calculation of the X-ray spectrum of the gas in the sensitivity bands of the satellites is required. The extent to which the dark matter presence and distribution, the SNIa's rate, and the cooling of the gas determine the past flow evolution – and then its present temperature and density profiles – can be derived only through numerical simulations.

## 4. Numerical simulations for an ISM origin of the VSC

In this section we explore the gas flow evolution in the parameter space ($R$, $\beta$, $\vartheta_{\rm SN}$), to find whether there are regions in which gas flows have the observed VSC characteristics at the present epoch. To this purpose we construct hydrodynamical evolutionary sequences specific to the galaxies in Table 1, NGC4697 and NGC4365; these have been chosen because, in Group 1, they have respectively the lowest and one of the highest values for $\sigma_*$, on which $L_{X,gas}$ and $T_{gas}$ mostly depend (§3). To tailor the model structures to these galaxies, we adopt also the observed values of the blue luminosities (Table 1) and of the core radii (King 1978). Initially only the two fundamental parameters $R$ (dark-to-light mass ratio) and $\vartheta_{\rm SN}$ (SNIa's rate) vary, while the dark-to-luminous core radii ratio $\beta = 2$; we explore later the effect of a variation of $\beta$. The mass profiles of such galaxy models are shown in Fig. 5, for $\beta = 2$ and $\beta = 10$.

The time-dependent equations of hydrodynamics with source terms, and the numerical code to solve them, are described in CDPR. The cooling curve is an analytical fit to the results given by the Raymond code for the thermal emission of a hot, optically thin plasma, at the collisional ionization equilibrium, and with solar abundance (Raymond and Smith 1977). The gas flow evolution is followed for 15 Gyrs.

The characteristics of the flows (X-ray luminosities in the *Einstein* and *ROSAT* sensitivity bands, spectral energy distributions, and X-ray surface brightness profiles) are compared directly to the data, after they are transformed into predicted "observed" quantities, as follows. Using the Raymond code, the spectral energy distribution in the bands of sensitivity of the satellites is computed in each one of the 100 grid points used for the hydrodynamical simulation. After projection, the energy distribution of the photon flux is corrected for the interstellar absorption along the line of sight, using the photo-absorption cross sections of Morrison and McCammon 1983. Finally the model spectral distribution is convolved with the *Einstein* IPC (Harnden *et al.* 1984) or *ROSAT* PSPC spectral response and effective area, to obtain the predicted counts distribution. The model X-ray surface brightness profile is convolved with the energy dependent PSPC point spread function (Hasinger *et al.* 1992). Such predicted models are compared directly to the observational data, and the results of the comparison are described below, in §4.1.2, 4.2.2, and §6.

### 4.1 NGC4697

#### 4.1.1 The X-ray Luminosity

In Fig. 6 we show how the various gas flow phases, at the end of the simulation, populate the $(R, \vartheta_{\rm SN})$ plane; the regions where $L_{X,gas}(15 \text{ Gyrs}) \sim L_{X,VSC}$ are also shown. Due to the low $\sigma_*$ value, a large region of the plane is occupied by global winds; to obtain $\chi(15\text{Gyrs})\approx 1$, that is to expect an outflow at the present epoch, we need a dark-to-light mass ratio $R \approx 40$, if $\vartheta_{\rm SN} = 1$.

For $R > 6$ the gas flow evolution is the "classical" one: wind/outflow/inflow; as anticipated in §3.2, $L_{X,gas}(15 \text{ Gyrs}) \approx L_{X,VSC}$ during the outflow. The $\vartheta_{\rm SN}$ value for this is very critical, because the flow is very sensitive to the global energetic balance. As an example, for ($R=9$, $\vartheta_{\rm SN}=0.38$) the gas is in outflow, and its $L_X$ can vary by a factor of 8 with $\vartheta_{\rm SN}$ varying by only 1%.

If $R\lesssim 6$ the flow doesn't follow the evolutionary history wind/outflow/inflow. If $\vartheta_{\rm SN}\lesssim 0.3$, i.e. excluding the cases when a global wind lasts up to the present epoch, a cooling catastrophe takes place in the first few Gyrs (Fig. 7) in the central regions of the galaxy, while the gas escapes the external regions at a high velocity up to the present, and we have a PW. This is due to the SNIa's heating being too low to prevent an inflow in the gas dense and tightly bound central regions, while it is high enough to drive out the gas in the external regions. The tendency is however to evolve into a subsonic outflow, and then eventually into an inflow, even for the outer regions, as the specific heating secularly declines (§3.1).

The flow behavior across the $(R, \vartheta_{\rm SN})$ plane can be explained looking at how the structure of the models changes across it. In the central regions of the galaxy the amount of luminous matter is always larger than that of the dark matter (see Fig. 5); increasing $R$, we bring to inner radii the dominance of the dark matter over the luminous one, but even for $R=9$, we have $M_h(<r)/M_*(<r) > 1$ only for $r > 11r_{c*}$. So, the central regions are always dominated (from a gravitational point of view) by the stellar structure, which is fixed across the plane. At low values of $R$, to avoid winds we decrease $\vartheta_{\rm SN}$, until it reaches a critical value $\vartheta_{PW}^c$ which is too low to prevent an inflow in the central regions, and we have a partial wind. At high values of $R$, $\vartheta_{PW}^c$ cannot prevent a global inflow, because also the external regions become very tightly bound.

The above results are sensitive to variations of the parameter $\beta$ especially during the outflow phase: a larger value for $\beta$ – decreasing the dark matter concentration: $M_h(<r)/M_*(<r) > 1$ for $r > 21r_{c*}$, if $R = 9$ and $\beta = 10$ (Fig. 5) – has the effect of retarding the onset of the global inflow. As a consequence, the values of $\vartheta_{\rm SN}$ corresponding to outflows should be slightly decreased in Fig. 6. In PW cases – when $R$ is low, and the luminous matter concentration is already high enough to cause a central inflow – a $\beta > 2$ has little influence. Values of $\beta$ smaller than $\sim 2$ are not plausible, since dark matter halos around galaxies (if present) seem to be diffuse (Mathews 1988; Bertin, Saglia, and Stiavelli 1992).

### 4.1.2 The X-ray Spectrum

The predicted pulseheight distributions of a few representative cases of gas flows having $L_{\mathrm{X},gas}(15 \text{ Gyrs}) \approx L_{X,VSC}$ are shown in Fig. 8. Here also the pulseheight distribution corresponding to the best fit temperature of the VSC from *Einstein* IPC data is shown, together with that corresponding to the 90% upper limit on $T_{VSC}$. The predicted distributions are closer to the best fit distribution for those models with low values of $\vartheta_{\mathrm{SN}}$ and $R$, i.e. PWs, when the gas temperature is lower. For $R \gtrsim 5$ the peak in the counts distribution of the gas flow models takes place at energies higher than in the best fit case, and it is clearly displaced for $R = 9$, $\vartheta_{\mathrm{SN}} = 0.38$. However, all the predicted distributions lie within that corresponding to $kT = 0.5$ keV, the 90% confidence upper limit on the VSC temperature, because the average X-ray emission temperature of gas flows is $kT = 0.2 - 0.3$ keV. In Fig. 9 we show the comparison of the average Group 1 spectrum with that of the ($R = 3$, $\vartheta_{\mathrm{SN}} = 0.2$) model, to which a hard thermal component has been added.

In conclusion, for a galaxy with the optical characteristics of NGC4697, an ISM origin of the VSC is in principle possible; the strongest constraints in the parameter space come from the required $L_{\mathrm{X},gas}$ value, not from that of the gas emission temperature. In fact the statistical uncertainties in deriving the temperature of the VSC from *Einstein* IPC data are comparable to the spread of the model temperatures. So, these large uncertainties don't allow any discrimination between different values of $R$ and $\vartheta_{\mathrm{SN}}$.

### 4.2 NGC4365

This galaxy, belonging to the Virgo cluster, has been observed also by the *ROSAT* PSPC. In the modelling we try to reproduce the results of these more accurate observations.

### 4.2.1 The X-ray Luminosity

In Fig. 10 we show the ($R$, $\vartheta_{\mathrm{SN}}$) plane for NGC4365. The main features of the gas flow behavior are the same as in the case of NGC4697; here we point out the main differences, and some characteristics of the gas flows which have been investigated in more detail, since more accurate observations are available.

Due to the larger value of $\sigma_*$, a larger region of the ($R$, $\vartheta_{\mathrm{SN}}$) plane is occupied by global inflows, and even with $R \sim 0$ the flows can be luminous enough (or even too luminous); the $\vartheta^c_{PW}$ values are higher than for NGC4697; the models with $L_{\mathrm{X},gas}(15 \text{ Gyrs}) \sim L_{X,VSC}$ are in PW up to $R \approx 5$. For $3 \lesssim R < 5$ the global wind phase is actually followed by a global outflow phase, but the emission of the latter is never sufficiently high: a central cooling catastrophe develops, originating a PW, *before* the outflow phase can reach $L_{X,VSC}$ (Fig. 11). Global outflows can have $L_{X,gas} \gtrsim 6 \times 10^{39}$ erg s$^{-1}$ only for $R \gtrsim 5$. The $\vartheta_{SN}$ value to have $L_{X,gas}$(15 Gyrs)$\sim L_{X,VSC}$ is specially critical if $R = 5 - 6$, the lowest $R$ values for which the outflow can be luminous enough: enough gas mass has accumulated within the galaxy only just before the central cooling catastrophe occurs (see the $R = 5$ case in Fig. 11).

As for NGC4697, changing the value of $\beta$ has some effect only during outflows; for instance, if $R = 9$ and we increase $\beta$ from 2 to 10, $\vartheta_{SN}$ goes from 0.68 to 0.59; if $R = 6$, increasing $\beta$ from 2 to 20 lowers $\vartheta_{SN}$ from 0.55 to 0.43.

### 4.2.2 The X-ray Spectrum

In Table 3 we show the average emission temperatures for a few representative models having $L_{X,gas}$(15 Gyrs) $\sim L_{X,VSC}$ in the *ROSAT* band; the emission region considered is the same used by Fabbiano *et al.* 1994 to derive the observed spectral counts distribution (i.e. a circle of 250″ radius). Going from PWs to outflows, $kT$ varies from 0.4 to 0.7 keV.

In Fig. 12a we plot the observed counts distribution, together with the superposition of the spectrum of the gas flow – corresponding to a galaxy model with $R = 3$ and $\vartheta_{SN} = 0.35$ – and the spectrum of a bremsstrahlung component with $kT \sim 4$ keV, both convolved with the *ROSAT* response. The interstellar absorption is fixed at the line of sight value given in Table 1; the temperature and normalization of the hard component have been chosen so as to minimize the $\chi^2$ of the fit with the observed counts distribution. From Fig. 12a we see how difficult it is to reproduce the observed peak at energies around 0.2 keV; this is true for every galaxy model. The case in Fig. 12a can be brought into much better agreement with the observational data if a column density $N_H = 3 \times 10^{19}$ cm$^{-2}$ is assumed (Fig. 12b).

In conclusion, the gas spectrum is not as soft as that of the VSC, and cannot explain the observed counts distribution at low energies, under reasonable conditions, even when no dark matter is present, and with a very low SNIa rate (see the $R = 0$ and $\vartheta_{SN} = 0.18$ case in Table 3), because of the value of $\sigma_*$.

### 4.3 Conclusions

The numerical simulations we have performed, tailored precisely on a couple of galaxies in which the VSC has been measured, show that hot gas flows can easily be characterized by a luminosity $L_{X,gas} \sim L_{X,VSC}$, for acceptable values (within the present uncertainties) of the input parameters describing the amount and distribution of dark matter, and the SNIa explosion rate. The analysis of the X-ray spectra of gas flows reveals instead that these flows can be responsible for the VSC only in those galaxies having a low value of $\sigma_*$ ($\lesssim 200$ km s$^{-1}$), such as NGC4697. Among these galaxies, the requirement $L_{X,gas} \sim L_{X,VSC}$ implies a low value of $\vartheta_{SN}$ ($\lesssim 0.4$ if $R \lesssim 10$).

*ROSAT* observations, confirming the existence of a VSC, indicate that its temperature doesn't depend on the $\sigma_*$ value of the host galaxy: at 99% confidence level $kT_{VSC} < 0.3$ keV in NGC4365, whose $\sigma_*$ value prevents the average gas emission temperature from falling below 0.4 keV. This fact leads us to conclude that by itself a hot ISM cannot generally explain the VSC.

## 5. Stellar Sources as Origin of the VSC

In the following we explore the alternative possibility that discrete sources have the required spectral characteristics and luminosity to account for the very soft emission in early-type galaxies. *ROSAT* has observed supersoft sources in M31 and the Magellanic Clouds, and the presence of a soft component in RS CVn systems has been well established, while late type stellar coronae have been known to be sources of very soft X-ray emission since the *Einstein* observations.

### 5.1 Supersoft Sources

*ROSAT* observations, during the All-Sky Survey and deep pointings, have revealed strong supersoft X-ray emission in several stellar sources; in a fraction of them such emission had been revealed previously by *Einstein* observations. In the Large Magellanic Cloud (LMC) the supersoft sources are CAL 83 and CAL 87, optically identified with close binary systems, and RXJ 0527.8-6954 (Greiner *et al.* 1991), and the variable RXJ0513.9-6951 (Schaeidt *et al.* 1993). In the Small Magellanic Cloud (SMC) supersoft sources are PN67 (at the location of a planetary nebula), 1E0035.4-7230, and SMC3, a symbiotic star (Wang and Wu 1992; Kahabka and Pietsch 1992). In our Galaxy *ROSAT* detected GQ Mus, a recent nova (Ögelman *et al.* 1993). These are among the softest X-ray sources observed, and it has been suggested that they form a separate new class of X-ray objects, not previously encountered in our Galaxy because of the interstellar extinction (Schaeidt *et al.* 1993). Their similar spectra – the bulk of the X-ray emission is below 0.5 keV – can be described by black body radiation with $kT_{bb} = 20 - 50$ eV and X-ray luminosity from a few$\times 10^{37}$ erg s$^{-1}$ up to $\sim 10^{38}$ erg s$^{-1}$. Finally, 15 supersoft sources have also been discovered in M31 (G. Hasinger 1993, private communication).

The origin of the soft spectra and the nature of the compact object in such supersoft sources are still unclear: they could be neutron stars accreting matter at or above the Eddington rate (Greiner *et al.* 1991), or they might represent a long sought class of white dwarfs (WDs) burning accreted matter (van den Heuvel *et al.* 1992). This second interpretation seems more likely, thanks also to the nature of the accreting object optically identified in a few cases (see for example the case of SMC3 or that of the nova GQ Mus). In the van den Heuvel *et al.* model for CAL83 and CAL87 the X-ray emission is produced by steady nuclear burning on the surface of a $0.7 \div 1.2$ $M_\odot$ WD; the accreted matter is not ejected in thermonuclear nova flashes, thanks to the high accretion rate $[(1-4)\times 10^{-7} M_\odot \text{yr}^{-1}]$.

### 5.1.1 Can Supersoft Sources Be the VSC Origin?

We examine here whether the presence of supersoft sources can explain the X-ray spectra of those early-type galaxies where the VSC was discovered. Fitting the average Group 1 X-ray spectrum with a black body component of $kT_{bb} = 50$ eV, instead of a 0.2 keV thermal component, the goodness of the fit remains the same; $kT_{bb} = 150$ eV at the best fit. The contribution of the black body component is approximately half of the total flux in the (0.2–4) keV band, and so, for example, $\approx 500$ supersoft sources are needed to explain the soft emission of NGC4697.

Fitting the X-ray spectrum of NGC4365 with a black body component of $kT_{bb} \sim 50$ eV, in place of a very soft thermal component, the fit gets worse: if we fix $N_H$ at the line of sight value, the minimum $\chi^2$ increases from 24.9 (with 18 degrees of freedom) to 37.8 (with 19 degrees of freedom). Allowing $kT_{bb}$ to vary, at the best fit $kT_{bb} \sim 110$ eV, and still the fit is slightly worse than with a 0.2 keV thermal component (minimum $\chi^2 = 26.7$ with 18 degrees of freedom). In Fig. 13 (which can be compared to Fig. 4a) the situation is summarized. If $kT_{bb} \sim 110$ eV, $L_{X,bb} \sim 3 \times 10^{40}$ erg s$^{-1}$, and so we need $\approx 300$ supersoft sources to account for the soft emission of NGC4365.

In conclusion, the supersoft sources can plausibly account only for a fraction of the very soft emission of early-type galaxies, because they are slightly too soft. Also, something more needs to be discovered concerning their nature (stellar components, origin and duration of the soft phase, variability) to investigate whether their collective emission can be significant in early-type galaxies. For example, the interpretation model of van den Heuvel *et al.* 1992 requires as donor stars respectively a normal F-G star with mass $\sim 1.4 - 1.5\ M_\odot$ for CAL 87, and either an evolved main sequence star of $\sim 1.9 - 2.M_\odot$, or a post main sequence subgiant of lower mass ($M \gtrsim 1.5 M_\odot$) for CAL 83. Such companions are unlikely to be present in the old population of an elliptical galaxy. In other cases the estimated companion mass is much lower [see, e.g., the case of GQ Mus, a classical nova, where the mass of the secondary is $< 0.2\ M_\odot$, Ögelman *et al.* 1993]. Recently, Rappaport *et al.* (1994) estimate that in our Galaxy today there should be more than 1000 systems with the characteristics required by the van den Heuvel *et al.* model.

Finally, it is interesting to note that were the supersoft sources WDs burning accreted matter, they would represent a long sought class of sources, which could also be the progenitors of type Ia supernovae (Munari and Renzini 1992, Kenyon *et al.* 1993). In this case the accreting WD would explode after having grown to the Chandrasekhar limit, or even before (helium detonation SNIa's). We estimate here what SNIa rate of explosion ($R_{exp}$) is expected if $L_{X,VSC}$ comes entirely from WDs accreting matter and then exploding as SNIa's: $R_{exp} \approx 10^{-52} L_{X,VSC}/\alpha M_{acc}$ s$^{-1}$, where $\alpha$ is the fraction of the bolometric luminosity produced by the hydrogen burning, collected by the *ROSAT* satellite; $M_{acc}$ is the mass which has to be accreted before explosion, in solar masses; and the energy yield per unit mass by hydrogen burning, for 0.7 hydrogen mass fraction, is $\sim 4.5 \times 10^{18}$ erg g$^{-1}$. On average, the progenitor WD has to accrete $\sim 0.2$ $M_\odot$ before exploding (Munari and Renzini 1992), or even less if it undergoes helium detonation (Woosley and Weaver 1993); so $R_{exp} \approx 1.5 \times 10^{-11}/\alpha$ s$^{-1}$ in NGC4365. Since the observed rate is $R_{SN} \approx 4.5 \times 10^{-10} \vartheta_{SN}$ s$^{-1}$ for this galaxy (from §3.1), to reproduce it we need $\alpha \lesssim 0.03/\vartheta_{SN}$. Note that $\alpha = 0.01, 0.24, 0.82$ for $kT_{bb} = 10, 20, 50$ eV, and so most of $L_{X,VSC}$ should come from objects with quite low temperatures. This requires to make another check, i.e. whether the UV flux that would be produced is consistent with that observed. Greggio and Renzini (1990) derive for ellipticals $L_{UV}/L_{BOL} < 0.02$; imposing this constraint on the UV emission of supersoft sources, we derive $\alpha > 2 \times 10^{-3}$ in NGC4365. So, the possible link between $L_{X,VSC}$, supersoft sources, and SNIa's is not ruled out by current observational constraints.

*5.2 Late Type Stars*

In comparison to other types of stellar X-ray sources which are powered by accretion, normal stars are very faint sources of X-ray emission ($L_X \approx 10^{26} - 10^{30}$ erg s$^{-1}$). Their collective emission can still be substantial, because dwarfs of all spectral types from F to M are sources of X-ray emission (e.g. Schmitt *et al.* 1990); particularly, it is interesting to consider the X-ray properties of K and M dwarfs. These stars are the most numerous in a single burst old stellar population like that of elliptical galaxies. In such stars the X-ray emission is thermal, and originates in the stellar coronae. From a systematic investigation of *Einstein* IPC data, Schmitt *et al.* 1990 find that a two-temperature description is always required for the coronae of K and M dwarfs when high signal-to-noise spectra are available. The values of these temperatures cluster around $kT \sim 1.4$ keV and $kT \sim 0.2$ keV.

We can estimate the collective emission due to the lower temperature component of K and M dwarfs in ellipticals by multiplying their expected number ($N_K$ and $N_M$ respectively) by their average X-ray luminosity ($<L_X>_K$ and $<L_X>_M$). Assuming that K and M dwarfs were generated with the Salpeter initial mass function, their expected number in a single burst stellar population is:

$$N_K = \int_{0.6}^{0.85} A \cdot M^{-2.35} dM \qquad (1)$$

and

$$N_M = \int_{0.08}^{0.6} A \cdot M^{-2.35} dM \qquad (2)$$

where $A = 2.9 L_B$ (CDPR), $M = 0.85 M_\odot$ is the turn-off mass after 15 Gyrs, $M_{inf} = 0.08 M_\odot$. So we derive $N_M = 60.7 L_B$ and $N_K = 1.61 L_B$, with $L_B$ in $L_\odot$; in the following we consider just the M dwarf contribution, since $<L_X>_M$ and $<L_X>_K$ are similar. Adopting $<L_X>_M = 2 \times 10^{28}$ erg s$^{-1}$ (Rosner *et al.* 1981), we obtain $L_{X,M} = N_M <L>_M = 1.2 \times 10^{30} L_B$. This means $L_{X,M} \approx 2 \times 10^{41}$ erg s$^{-1}$ for NGC4697, and $L_{X,M} \approx 8 \times 10^{40}$ erg s$^{-1}$ for NGC4365: even if just a fraction of $L_{X,M}$ is due to the 0.2 keV component, M dwarfs seem to be able to explain the VSC in early-type galaxies. In general, $\log(L_{X,M}/L_B) \approx 30$, while for Group 1 galaxies $\log(L_{X,VSC}/L_B) = 29.0 - 29.7$.

Two comments are however in order. The first is that in late-type stars the emission is strongly dependent on the rotation rate (Pallavicini *et al.* 1981), because coronal heating is strongly related to the presence of a magnetic dynamo (Rosner 1991). Probably an "age effect", reducing the angular momentum, also reduces the X-ray luminosity, and this is particularly crucial for population II M dwarfs. Barbera *et al.* 1993 have systematically investigated the presence of this effect: using all the available *Einstein* IPC data, they have identified a subsample representative of the population of M dwarfs in the solar neighborhood; using a kinematic criterion to determine the age, they have confirmed the decrease in X-ray luminosity going from young-disk to old-disk population. For old M dwarfs they find $< L_X >_M = (0.8 - 1) \times 10^{27}$ erg s$^{-1}$ (going from late to early M). We derive then $\log(L_{X,M}^{\text{old}}/L_B) \approx 28.7$; for NGC4365 $L_{X,M}^{\text{old}} \sim 3.5 \times 10^{39}$ erg s$^{-1}$, while $L_{X,VSC} \sim 2.7 \times 10^{40}$ erg s$^{-1}$. So the age effect probably rules out the M dwarfs as the main origin of the VSC in ellipticals; the contribution of these stars could even be much lower, when we consider that the *ROSAT* pointed observation of an old M dwarf such as the Barnard star shows that its $L_X = 3 \times 10^{25}$ erg s$^{-1}$ (T. Fleming 1993, private communication).

The second consideration is that the Salpeter initial mass function probably doesn't hold down to stellar masses as low as 0.08 $M_\odot$. Tinney *et al.* 1992 find that in the solar neighborhood the initial mass function clearly flattens for stellar masses $< 0.2 M_\odot$. If this is the case also in early-type galaxies, we have overestimated $N_M$. Unfortunately we don't really know what the IMF is in ellipticals; we can just note that the adopted Salpeter IMF would imply a mass-to-light ratio of $M_*/L_B \approx 13$, within the range of values recently estimated by Guzmán, Lucey, and Bower 1993 ($M_*/L_B = 8 \div 25$, assuming $L_V = 1.67 L_B$ as Greggio and Renzini 1990). If we want to explain the VSC with old M dwarfs, we have to increase $N_M$ by at least of a factor of 4 (8 in NGC4365). This implies an IMF steeper than the Salpeter function, with slope $> 2.6$ in eq. (2), which in turn implies $M_*/L_B > 19$.

In conclusion, late type stellar coronae do have the required spectral characteristics to explain the VSC in ellipticals, but due to an age effect we expect their collective emission to be too low to account for it all, unless the IMF for M dwarfs is much steeper than the Salpeter function, and hence is at variance with the IMF in the solar neighborhood.

*5.3 RS CVn systems*

RS Canum Venaticorum (RS CVn) systems are chromospherically active objects, typically consisting of a G or K giant or subgiant, with a late type main sequence or subgiant companion (Linsky 1984). As *Einstein* observations have shown, RS CVn's are the most X-ray luminous late type stars ($L_X \sim 10^{29} - 10^{31.5}$ erg s$^{-1}$), and their quiescent coronae can be well modeled by a thermal plasma with two temperatures (see e.g. Schmitt *et al.* 1990). Recently, Dempsey *et al.* 1993 presented the results of their analysis of X-ray spectra of 44 RS CVn systems obtained during the *ROSAT* All-Sky Survey with the PSPC. They find that a bimodal distribution of temperatures centered near 0.2 and 1.4 keV fit the data best.

We examine here the soft X-ray emission expected from RS CVn systems in early type galaxies. Though most of Dempsey *et al.* 's systems belong to population I, we take $4 \times 10^{30}$ erg s$^{-1}$ as an average value for the soft X-ray luminosity of a system (Dempsey *et al.* 's Fig. 3). The spatial density of RS CVn's has been estimated by Ottmann and Schmitt 1990 for the Galaxy; here we estimate it for an old stellar population as follows. The number of RS CVn is a fraction of the number of G or K giants and subgiants (respectively $N_{giant}$ and $N_{subg}$); $N_{giant}$ and $N_{subg}$ can be calculated from the knowledge of the time spent by stars in such evolutionary phases (see e.g. Renzini 1989). Assuming that low mass metal rich stars spend approximately 1 Gyr on the subgiant branch, and 1 Gyr on the red giant branch, we have $N_{subg} + N_{giant} \approx 0.1 L_{\rm B}$. Taking half this number for those in binary systems, and another factor ($a$) for those which become RS CVn, we derive the collective contribution $L_{X,RSCVn} = 4 \times 10^{30}(N_{subg} + N_{giant}) \approx 2 \times 10^{29} a L_{\rm B}$; or log $(L_{X,RSCVn}/L_{\rm B}) = 28.6$ if $a = 0.2$. So we conclude that RS CVn systems represent another class of likely contributors to the VSC in early-type galaxies, especially thanks to their spectral characteristics, but they are certainly not the unique explanation of the phenomenon.

## 6. A Model for the X-ray Emission of NGC4365

We propose here a model for the X-ray emission of NGC4365, to explain the results obtained from the *ROSAT* pointed observation of this galaxy (Fabbiano *et al.* 1994). The observed X-ray spectrum and surface brightness profile ($\Sigma_X(R)$) allow us to make some conjectures about the dynamical phase of the gas flow, and the different components of the X-ray emission. From the resulting model we also obtain information on the dark matter content of this galaxy, its distribution, and the SNIa rate.

### *6.1 The X-ray surface brightness profile*

$\Sigma_X(R)$ obtained by *ROSAT* is shown in Fig. 14. Assuming that the hard component present in the X-ray spectrum (§2) is due to low-mass accreting binaries, its spatial distribution probably follows that of the stars; the same holds for the VSC, if it comes from stellar sources. In Fig. 14 we also plot the superposition of two brightness profiles, having each the spectral energy distribution and the normalization of the two components derived from the best fit to the observed spectrum; their spatial distribution is the same as that of the optically luminous matter (King 1978), and has been convolved with the PSPC response. Taken at face value, the observed $\Sigma_X(R)$ appears less peaked than that of this model. However, the discrepancy is at the center, where eventual uncertainties on the PSF are more crucial. The way to really solve the point is to observe the galaxy with the ROSAT HRI, which we will propose to do. For the purpose of this paper, we will take this discrepancy at face value, and look for possible explanations. This exercise demonstrates the complexity of solutions one may have.

The first way out comes from assuming that some cold material is present at the center of the galaxy to absorb the X-ray radiation (see, e.g., White *et al.* 1991). In this case, the 'observed' $\Sigma_X(R)$ could look like that in Fig. 15, where the absorbing material is confined within a spatial radius of 4 kpc, and produces a column density of $\sim 5 \times 10^{21}$ cm$^{-2}$ (the spectral counts distribution is that observed by construction). The required cold gas mass would be $M_{HI} \approx 2 \times 10^9$ $M_\odot$; from observations (Knapp *et al.* 1985) an upper limit of $1.5 \times 10^8$ $M_\odot$ for the HI mass has been estimated. Based on these results, we consider the hypothesis of an intrinsic absorption unlikely.

### *6.2 X-ray Emission Sources in NGC4365*

A hot ISM doesn't need to be distributed as the stars; indeed, its $\Sigma_X(R)$ can be very flat, during the outflow phase (Ciotti *et al.* 1992). So, introducing the presence of hot gas (as discussed earlier, this gas would be hotter the the observed VCS, and therefore its presence is not required by the X-ray spectral analysis) could help reproduce the particular shape of the observed $\Sigma_X(R)$. The amount and distribution of dark matter then play an important role, because – as simulations show – the gas density profile is flatter for larger and more extended (large $\beta$ values) dark matter halos.

The gas flow that allows the best agreement with both the observed X-ray spectrum and $\Sigma_X(R)$ must have the following characteristics: its $\Sigma_X(R)$ is mostly flat, and its $L_X$ is not larger than $\sim 2 \times 10^{40}$ erg s$^{-1}$, since a higher value cannot be reconciled with the observed spectrum. To reproduce the latter at high ($> 1$ keV) and low ($< 0.5$ keV) energies, two further components must be added to the spectrum of the gas flow: a hard component, and a very soft component. The temperature and normalization of these additional components are determined by a best fit to the observed spectrum.

After an accurate exploration of the $(R,\beta,\vartheta_{\rm SN})$ parameter space, we found that the galaxy model best reproducing the data has ($R = 9$, $\beta = 9.4$, $\vartheta_{\rm SN} = 0.59$). The X-ray properties of the three components are summarized in Table 4. $L_{X,VSC}$ strongly depends on the assumed $N_H$ value; here the value in Table 1 is assumed. The very soft emission can also be produced by a black body component of $kT_{bb} = 50$ eV, rather than by a Raymond thermal component, and so the VSC temperature in this model is close to that of the supersoft sources (§5.1). To reproduce the whole $L_{X,VSC}$ in this way we need $\sim 200$ supersoft sources.

The resulting "best fit" spectrum is plotted in Fig. 16; $\chi^2/\nu = 1.6$, where $\nu = 18$ is the number of degrees of freedom (the lower and higher energy spectral bins have been excluded from the fit, see e.g. Fabbiano *et al.* 1994). The $\Sigma_X(R)$ resulting from the superposition of the three components is plotted in Fig. 17 ($\chi^2/\nu = 1.3$, for $\nu = 11$); the spatial distribution of the 0.1 keV and the 5 keV components is the same as that of the stars.

In conclusion, we suggest for this galaxy a SNIa's explosion rate of $\approx 0.6$ the Tammann 1982 rate, and a mass model in which there is a considerable amount of dark matter, distributed mostly at large radii. For the reasons explained at §6.1, we stress that this result apply insofar the data support it, and needs more observational confirmation, such as can be obtained from the ROSAT HRI. The stellar velocity dispersion profile of NGC4365 is observed only within the central $\sim 30''$ (Bender and Surma 1992), so a useful check for the dark matter effects expected from our galaxy model is not possible. Bender and Surma 1992 also find that NGC4365 might have suffered a merging process in the past, based on the presence of a kinematically decoupled core component. If this is true, and a starburst followed the merger, the stellar population of NGC4365 might be peculiar.

## 7. Conclusions

In this paper we have investigated possible origins for the very soft X-ray emission ($kT \sim 0.2$ keV) recently discovered in low $L_X/L_B$ early-type galaxies. We considered both an origin in gas flows, performing numerical simulations of their behavior tailored to two galaxies (NGC4697 and NGC4365), and an origin in stellar sources. Our findings are as follows:

1) The X-ray luminosity of gas flows can be comparable to that of the very soft emission during the outflow or the partial wind dynamical phases. Both phases can be found at the present epoch only if the SNIa rate of explosion is lower than the standard (Tammann 1982) rate. The PW phase occurs only for values of the dark/luminous matter ratio up to 5–6, and the outflow for larger values.

The average emission temperature of gas flows is comparable to that of the very soft emission only in galaxies with quite shallow potential wells, i.e. with low values of the central stellar velocity dispersion ($\sigma_* \lesssim 200$ km s$^{-1}$). Due to the larger $\sigma_*$ value, the average emission temperature of gas flows in NGC4365 ($kT > 0.4$ keV) is higher than that of the very soft emission ($kT < 0.3$ keV at 99% confidence level) determined from the *ROSAT* pointed observation of NGC4365. So, by itself a hot ISM cannot generally explain the very soft emission.

2) Concerning stellar sources, we investigated the possible contributions to the very soft emission coming from supersoft sources, late type stellar coronae, and RS CVn systems.

The typical emission temperature of supersoft sources ($kT_{bb} < 70$ eV) is too low to explain entirely the very soft emission of early-type galaxies, although these objects might make an important contribution. In principle, if their nature is that of accreting white dwarfs, they could also account for the observed SNIa explosion rate.

M dwarf stars' coronae possess a thermal component at a temperature close to that of the very soft emission of early-type galaxies, but they are not likely to account entirely for it, because an age effect which reduces their X-ray luminosities is expected in old stellar populations. The initial mass function of M dwarfs should be much steeper than the Salpeter IMF to compensate for the age effect, but this may conflict with the observed $M_*/L_B$ ratios of ellipticals.

RS CVn systems can also give a contribution, being the most X-ray luminous late type stellar sources, and possessing a soft ($kT \approx 0.2$ keV) thermal component.

3) We construct a detailed model for the X-ray emission of NGC4365, which can account for the results of the *ROSAT* PSPC pointed observation. We suggest the presence of three contributors, approximately equally important: evolved stellar sources such as low-mass X-ray binaries, producing the hard emission (energies $> 1$ keV); a hot ISM, to better reproduce the X-ray surface brightness profile ($kT_{gas} \sim 0.6$ keV); stellar sources with very soft emission ($kT \approx 0.1$ keV). A major fraction of the latter could come from supersoft sources such as those in 2). The galaxy model that produces a gas flow with the required properties has a dark-to-luminous mass ratio of 9, with the dark matter very broadly distributed, and a SNIa explosion rate of $\sim 0.6$ the Tammann rate. This galaxy model can be confirmed with ROSAT HRI pointed observations, which will give accurate measurements of the central $\Sigma_X$.

4) In NGC4697, due to the low value of the central stellar velocity dispersion ($\sigma_* = 186$ km s$^{-1}$), the ISM is in a wind regime in most of the parameter space describing the dark matter presence and the supernova explosion rate. If we don't assume that considerable variations from galaxy to galaxy exist in the amount of dark matter and SNIa rate, the very soft emission should come entirely from stellar sources in NGC4697. This can be tested by future *ROSAT* pointed observations of this galaxy.

### Acknowledgements

We are grateful to A. D'Ercole for the hydrodynamical routines, F. Fiore for enlightments on the instrumental responses, D.W. Kim for the Group 1 X-ray spectrum, and J. Raymond for having kindly provided his thermal X-ray emission code; we thank L. Ciotti, P. Eskridge, T. Fleming, J. Orszak and M. Birkinshaw for discussions and comments, and A. Renzini for useful suggestions. We finally thank the referee, G. Hasinger, for useful comments. This work was supported by a Smithsonian Institution pre-doctoral fellowship, and by NASA LTSA grant NAGW-2681.

FIGURE CAPTIONS

**Figure 1**: The $L_X$–$L_B$ diagram for early-type galaxies; data (detections only) are from Fabbiano, Kim, and Trinchieri (1992). The dashed line is the luminosity expected from SNIa heating $L_{SN}$, assuming the Tammann (1982) rate of explosion. Group 1 galaxies are indicated with different symbols (full squares).

**Figure 2**: The average spectral energy distribution of Group 1 galaxies, obtained by the *Einstein* IPC (open squares with errorbars). Also plotted are the two thermal components which best fit this distribution (dotted and dashed lines), and their superposition (solid line).

**Figure 3**: 68% and 90% probability confidence contours for the temperatures (in keV) of the two components which best fit the average spectrum of Group 1 galaxies: the VSC (abscissa) and the hard component (ordinate). The increments above the $\chi^2_{min}$ value correspond to 3 interesting parameters, as explained in the text (Avni 1976).

**Figure 4**: a) as in Fig. 3, for the *ROSAT* PSPC spectral data of NGC4365; also the 99% probability confidence contour is shown. b) the *ROSAT* PSPC X-ray spectrum of NGC4365, together with the VSC (dotted line), the hard component (dashed line), and their superposition (solid line), at the best fit (from Fabbiano *et al.* 1994).

**Figure 5**: The radial trend of the dark to luminous mass ratio $M_h(<r)/M_*(<r)$, for the cases $R = 1$ (lower lines) and $R = 10$ (upper lines); $\beta = 2$ for the dashed lines, and $\beta = 10$ for the dot-dashed lines.

**Figure 6**: The $(R, \vartheta_{SN})$ plane for NGC4697. The dashed lines enclose the region of PWs. The solid lines enclose the region where the models have $L_{X,gas} \sim L_{X,VSC}$; they are coincident in the outflow regime, when a small variation of $\vartheta_{SN}$ causes a large variation of $L_{X,gas}$.

**Figure 7**: The time evolution of the fundamental hydrodynamical quantities describing the gas properties: velocity (v<0 for inflows, v>0 for outflows), temperature T and number density n. Plotted is the PW case ($R = 5$, $\vartheta_{SN} = 0.27$) for NGC4697. Dashed lines refer to $t$=0.75 Gyrs, dotted lines to $t$=4.5 Gyrs, and solid lines to the present epoch ($t$=15 Gyrs).

**Figure 8**: The X-ray spectrum of the VSC from the analysis of *Einstein* spectral data (solid line), and that of a thermal component with $kT = 0.5$ keV, the 90% confidence upper limit on the VSC temperature (dotted line). The other lines are the gas spectra (convolved with the IPC response) for the following models for NGC4697: $R = 9$, $\vartheta_{SN} = 0.38$ (dashed line); $R = 6$, $\vartheta_{SN}$=0.28 (long dashed line); $R = 4$, $\vartheta_{SN} = 0.23$ (dot-dashed line); $R = 3$, $\vartheta_{SN}$=0.2 (dot-long dashed line). In order to compare the shapes, each spectrum has been normalized to the number of counts in the channel where it has the maximum.

**Figure 9**: The observed spectral count distribution for Group 1 galaxies (open squares with errorbars) compared with the superposition (solid line) of the convolved spectral distributions of the gas flow in NGC4697, for a model with $R = 3$,

$\beta = 2$ and $\vartheta_{\rm SN} = 0.2$ (dotted line), and of a hard ($kT \sim 5$ keV) thermal component (dashed line).

**Figure 10**: The ($R$, $\vartheta_{\rm SN}$) plane for NGC4365. The solid lines enclose the region where the models have $L_{X,gas} \sim L_{X,VSC}$, and the dashed lines the PW region.

**Figure 11**: The time evolution of the X-ray luminosity in the $ROSAT$ sensitivity band for the following models for NGC4365: $R=1$, $\vartheta_{\rm SN} = 0.25$ (solid line); $R=2$, $\vartheta_{\rm SN} = 0.3$ (dotted line); $R = 3.5$, $\vartheta_{\rm SN} = 0.36$ (dashed line); $R=9$, $\vartheta_{\rm SN} = 0.68$ (dot-dashed line). The three PWs can be recognized from the early cooling catastrophes (peaks in $L_{\rm X}$). To show the strong sensitivity to $\vartheta_{\rm SN}$ during the outflow phase, the case of $R = 5$ has also been plotted (long dashed lines): in order of increasing $L_{X,gas}$, the models have $\vartheta_{\rm SN} = 0.500, 0.498, 0.495$ (this last is an inflow at the present epoch). The parallel solid lines represent the range of $L_{X,VSC}$ values allowed by the data.

**Figure 12**: The observed spectral counts distribution for NGC4365 (open squares with errorbars) together with the superposition (solid line) of the distribution of a gas flow of the ($R = 3$, $\vartheta_{\rm SN} = 0.35$) model (dotted line), and that of a hard component with $kT = 5$ keV (dashed line), both convolved with the PSPC instrumental response. In a) $N_H = 1.6 \times 10^{20}$ cm$^{-2}$, and in b) $N_H = 3.2 \times 10^{19}$ cm$^{-2}$.

**Figure 13**: 68%, 90%, and 99% probability confidence contours for the temperatures of a hard thermal component (ordinate), and of a black body component (abscissa), derived fitting the ROSAT PSPC data of NGC4365 (see §5.1.1).

**Figure 14**: The observed X-ray surface brightness profile of NGC4365 (open squares with errorbars), from Fabbiano *et al.* 1994; unbinned PI channels 26–247 have been considered, to eliminate ghost effects. Also plotted are the surface brightness profiles of the hard (long dashed line) and soft (short dashed line) thermal components, derived from the best fit of the observed X-ray spectrum (§2), and their superposition (solid line). The thermal components are distributed as the optical profile of the galaxy (King 1978), and have been convolved with the PSPC response, appropriate for their respective spectral energy distribution.

**Figure 15**: The same as in figure 14, but an intrinsic absorption corresponding to $N_H = 5 \times 10^{21}$ cm$^{-2}$ has been applied within the central 30$''$ (see §6.1).

**Figure 16**: The observed spectral energy distribution of NGC4365 (open squares with errorbars), together with the three components of the "best fit" model (§6.2) after convolution with the PSPC response: the hot ISM (dotted line), the very soft component ($kT \approx 0.1$ keV, dot-dashed line), and the hard component ($kT = 5$ keV, dashed line); the solid line is their superposition.

**Figure 17**: The X-ray surface brightness profile (solid line) of the superposition of the three components which best fit the observed X-ray properties of NGC4365 (§6.2). Open squares are the observed counts, the long dashed line is the hard thermal component, the short dashed line is the very soft $kT \approx 0.1$ keV thermal component, and the dotted line is the hot gas.

TABLE 1

Observational Properties

| NAME | $L_B^a$ $(10^{11}L_\odot)$ | $\sigma_*^b$ (km s$^{-1}$) | d $^c$ (Mpc) | $N_H^d$ $(10^{20}$cm$^{-2})$ | log $L_X^e$ (erg s$^{-1}$) | $\frac{L_{X,VSC}}{L_X}$ $^f$ | $kT_{VSC}^f$ (keV) |
|---|---|---|---|---|---|---|---|
| NGC4697 | 1.95 | 186 | 37.4 | 2.1 | 40.97 | 0.15 − 0.50 | $\lesssim 0.5$ |
| NGC4365 | 0.65 | 262 | 27 | 1.6 | 40.74 | 0.3 − 0.7 | 0.14–0.21 |

$^a$ total blue luminosity, from Fabbiano, Kim, and Trinchieri 1992
$^b$ central stellar velocity dispersion, from Whitmore, McElroy, and Tonry 1985
$^c$ distance, from Fabbiano, Kim, and Trinchieri 1992
$^d$ Galactic hydrogen column density in the line of sight of the galaxies, from Stark *et al.* 1992
$^e$ NGC4697 X-ray luminosity is the estimate derived from *Einstein* data by Fabbiano, Kim, and Trinchieri 1992, in the (0.2–4) keV range; $L_X$ for NGC4365 is derived from *ROSAT* data by Fabbiano *et al.* 1994, in the (0.1–2.0) keV range. From *Einstein* data, Fabbiano, Kim, and Trinchieri 1992 derived log$L_X$=40.42 for NGC4365, in the (0.2–4) keV range, assuming 1 keV thermal emission.
$^f$ 90% confidence range (see §2); the Galactic absorption in the line of sight of the galaxies has been assumed.

TABLE 2

$\chi(15 \text{ Gyrs})$, with $\vartheta_{SN} = 1$

| | | $\sigma_* = 186$ km s$^{-1}$ | | | $\sigma_* = 262$ km s$^{-1}$ | |
|---|---|---|---|---|---|---|
| R | 0 | 5 | 10 | 0 | 5 | 10 |
| β | | | | | | |
| 1 | 0.12 | 0.36 | 0.60 | 0.25 | 0.71 | 1.18 |
| 6 | 0.12 | 0.31 | 0.49 | 0.25 | 0.61 | 0.98 |
| 11 | 0.12 | 0.29 | 0.46 | 0.25 | 0.58 | 0.92 |

TABLE 3

NGC4365: gas flows average emission temperatures
(t=15 Gyrs)

| R | $\vartheta_{SN}$ | flow [a] phase | $kT_{gas}$ (keV) |
|---|---|---|---|
| 9 | 0.68 | O | 0.7 |
| 5 | 0.498 | O | 0.5 |
| 3 | 0.35 | PW | 0.4 |
| 0 | 0.18 | PW | 0.4 |

[a] $O=$ outflow; $PW=$ partial wind

TABLE 4

X-ray emission components for NGC4365

|  | $kT$ (keV) | $L_X(0.1-2 \text{ keV})$ ($10^{40}$ erg s$^{-1}$) | $L_X/L_{X,tot}$ |
|---|---|---|---|
| soft sources | $\approx 0.1$ | 2.0 | 0.35 |
| hard sources | 5 | 2.1 | 0.37 |
| hot ISM | 0.6 | 1.6 | 0.28 |

Note: the X-ray luminosities, and the average X-ray emission temperature of the hot gas, are calculated considering the same region used by Fabbiano *et al.* 1993 (a circle of 250″ radius).